# Field Dependent Phase Diagram of the Quantum Spin Chain $(CH_3)_2NH_2CuCl_3$

M. B. Stone[*], W. Tian[†], T. P. Murphy[¶], S. E. Nagler[*], and D. G. Mandrus[*]

[*]*Condensed Matter Sciences Division, Oak Ridge National Laboratory, Oak Ridge, TN 37831, USA*
[†]*Department of Physics and Astronomy, The University of Tennessee, Knoxville, Tennessee 37996, USA*
[¶]*National High Magnetic Field Laboratory, Florida State University, Tallahassee, Florida 32310, USA*

**Abstract.** Although $(CH_3)_2NH_2CuCl_3$ (MCCL) was first examined in the 1930's [1], there are open questions regarding the magnetic dimensionality and nature of the magnetic properties. MCCL is proposed to be a $S=1/2$ alternating ferromagnetic-antiferromagnetic spin chain alternating along the crystalline $a$-axis [2,3]. Proposed ferromagnetic ($J_{FM}$ =1.3 meV) and antiferromagnetic ($J_{AFM}$ =1.1 meV) exchange constants make this system particularly interesting for experimental study. Because $J_{FM}$ and $J_{AFM}$ are nearly identical, the system should show competing behavior between $S=1/2$ (AFM) and $S=1$ (FM) effects. We report low temperature magnetic field dependent susceptibility, $\chi$, and specific heat, $C_p$, of MCCL. These provide an initial magnetic-field versus temperature phase diagram. A zero-field phase transition consistent with long range magnetic order is observed at $T=0.9$ K. The transition temperature can be reduced via application of a magnetic field. We also present comparisons to a FM/AFM dimer model that accounts for $\chi(T,H=0)$ and $C_p(H,T)$.

MCCL crystallizes in the monoclinic structure (space group I2/a) at room temperature with lattice constants $a = 11.97$ Å, $b = 8.6258$ Å and $c = 14.34$ Å, and $\beta = 97.47°$ [2]. Willett proposed that the magnetic properties of MCCL arose from $S=1/2$ $Cu^{2+}$ ions coupled via Cu-halide-Cu bridges to form magnetic chains along the $a$-axis. MCCL was classified as an alternating sign exchange chain based on measurements of bulk magnetic susceptibility and structure calculations [3]. MCCL occupies a unique niche in the world of low-dimensional magnets with particular interest regarding integer spin chains.

Single crystals were prepared through slow evaporation of stoichiometric aqueous solutions of $CuCl_2 \cdot 2(H_2O)$ and $(CH_3)_2NH_2Cl$ [2]. The field dependence of $C_p$ was measured using a commercial heat pulse calorimeter between $T=1.8$ K and $T=300$ K with applied magnetic fields up to $H=8$ T perpendicular to the $a$-axis. $\chi(T)$ and magnetization for $T \geq 0.25$ K with fields parallel to the $a$-axis were measured using the 18 Tesla superconducting magnet at the National High Magnetic Field Laboratory employing a split coil susceptometer and cantilever magnetometer. Magnetization measurements for $T \geq 1.8$ K were performed using a commercial SQUID magnetometer.

Figure 1(a) is the temperature dependent $C_p$ of MCCL measured at several applied magnetic fields. We plot $C_p/T$ versus $T^2$ on a semi-log scale to emphasize low-temperature differences as a function of applied field. A broad maximum at approximately $T = 4$ K is observed in the zero field $C_p$ curve. Such a feature is indicative of a low-dimensional gapped AFM. The maximum is gradually reduced with increasing magnetic field. Above $T=7$ K, $C_p(T)$ increases with field; however, below $T=7$ K, $C_p(T)$ first increases (up to $H \approx 3$ T for $T=2$ K), then decreases with further increasing the magnetic field. The field at which this transition in behavior occurs increases with temperature. As indicated below, this may be attributed to the competition between effective $S=1/2$ and $S=1$ effects in MCCL.

Prior studies compared $C_p(T)$ and $\chi(T)$ with several 1D models; however, these failed to obtain simultaneous agreement for both measurements [4,5,7]. We obtain fair agreement for a simultaneous fit of $C_p(H,T)$ and $\chi(T)$ using the model introduced in Ref. [6]. This describes MCCL as being composed of equal fractions of non-interacting FM and AFM $S=1/2$ dimers. The FM dimers have a zero-field degenerate triplet ground state with a gap to the singlet state, $\Delta_s$, and the AFM dimers are composed of a non-magnetic singlet ground state with a gap to a zero-field degenerate triplet state, $\Delta_t$. Including field dependent Zeeman splitting of the triplet states, we are able to simultaneously fit both $C_p(H,T)$ and $\chi(T)$. The zero field spin-gaps obtained from this are $\Delta_s =37(1)$ K and $\Delta_t =13.78(6)$ K. The phonon contribution to $C_p$ includes a $aT^3$ harmonic and a small $bT^5$ anharmonic contribution ($b/a*100 = 0.115(8)\%$); no electronic or nuclear contribution was included. Figure 1 depicts the fit results for both the magnetic susceptibility (b) and the $C_p$ (a). We also indicate the field independent non-magnetic contribution to the heat capacity as well as the individual zero-field contributions of the AFM and FM dimer terms. Although the AFM/FM isolated dimer model is able to reproduce the field dependence of the heat capacity measurements, there is less agreement below T≈10 K. This is most likely due to inter-dimer interactions not included in this calculation.

We identify crossovers in the magnetic ground state with torque magnetometry. Figure 1(c) depicts the derived field-temperature phase diagram. Figure 1(d), $\chi(T,H=0)$, shows a cusp at $T=0.9$ K followed by a rapid reduction of magnetic susceptibility, and is indicative of a transition to AFM long-range order (LRO). This was also observed in field-dependent torque magnetometery and $\chi(T,H\neq0)$, and in a recent $C_p(T)$ measurement [6]. The LRO phase is suppressed in a magnetic field of $H$≈0.5 T. We also find low-field and high-field transitions. These occur at $H$≈2 T and $H$≈14 T. We associate the high field transition with the fully FM state. This was previously observed in a $T=1.8$ magnetization measurement up to $H=32$ T [3]. We designate the region between $H$≈2 T and $H$≈14 T as an intermediate field regime, where MCCL has a finite magnetization. The origin of the low field phase transition is unclear, it may correspond to the alignment of canted spins under the influence of applied magnetic field or potentially indicate the closing of the spin-gap.

We make a successful comparison to an AFM/FM isolated dimer model for the $T>1.8$ K phase diagram; however, we caution using thermodynamic measurements to characterize the dimensionality of a magnetic system. Unusual magnetic field effects in low temperature $C_p$ and the phase diagram indicate that further study is necessary to characterize MCCL.

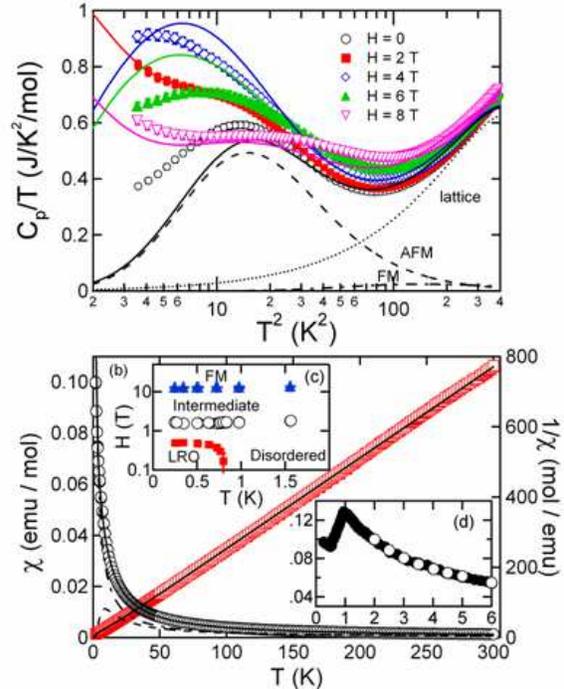

**FIGURE 1.** (a) $C_p(H,T)/T$ vs. $T^2$. (b) $\chi(T,H=500$ Oe) open circles left axis; $1/\chi$ open squares right axis. Solid line in (a) and (b) is dimer model discussed in text. Dashed (dot-dashed) line represents AFM(FM) portion for $H=0$. (c) Semi-log plot of $H$ vs. $T$ phase diagram. FM is fully polarized state. (d) Low-Temperature $\chi(T,H=0$ Oe) vs. $T$. Open symbols from SQUID magnetometer, closed symbols from AC susceptometer.

## ACKNOWLEDGMENTS

Oak Ridge National laboratory is managed by UT-Battelle, LLC, for DOE under DE-AC05-00OR22725. A portion of work performed at the National High Magnetic Field Laboratory, supported by NSF Cooperative Agreement No. DMR-0084173, the State of Florida and DOE. We acknowledge valuable discussions with M. W. Meisel.